\documentclass[12pt,preprint]{aastex}
 
\begin{document} 

\title{A Comprehensive Study of Pulse Profile Evolution in SGR 1806-20 \& 
SGR 1900+14 with the RXTE PCA} 

\author{Ersin {G\"o\u{g}\"u\c{s}}\altaffilmark{1,2}, 
Chryssa Kouveliotou\altaffilmark{2,3}, 
Peter~M.~Woods\altaffilmark{2,4},
Mark~H.~Finger\altaffilmark{2,4}, 
Michiel~van~der~Klis\altaffilmark{5}}
 
\altaffiltext{1}{Department of Physics, University of Alabama in 
Huntsville, Huntsville, AL 35899} 
\altaffiltext{2}{National Space Science and Technology Center, 320 Sparkman Dr.
Huntsville, AL 35805}
\altaffiltext{3}{NASA Marshall Space Flight Center, SD-50, Huntsville, 
AL 35812} 
\altaffiltext{4}{Universities Space Research Association} 
\altaffiltext{5}{Astronomical Institute ``Anton Pannekoek'' and CHEAF,
University of Amsterdam, 403 Kruislaan, 1098 SJ Amsterdam, NL}

\authoremail{Ersin.Gogus@msfc.nasa.gov} 

\begin{abstract} 

Soft Gamma Repeaters undergo pulse profile changes in connection with 
their burst activity. Here we present a comprehensive pulse profile history 
of SGR 1806-20 and SGR 1900+14 in three energy bands using Rossi X-ray 
Timing Explorer/Proportional Counter Array observations performed between 1996
and 2001. Using the Fourier harmonic powers of pulse profiles, we quantify 
the pulse shape evolution. Moreover, we determined the RMS pulsed count rates 
(PCRs) of each profile. We show that the pulse profiles of SGR 1806$-$20 
remain single pulsed showing only modest changes for most of our observing 
span, while those of SGR 1900+14 change remarkably in all energy bands. 
Highly significant pulsations from SGR 1900+14 following the 1998 August 27 
and 2001 April 18 bursts enabled us to study not only the decay of PCRs in 
different energy bands but also their correlations with each other.

\end{abstract} 


\section{Introduction} 

Soft gamma repeaters (SGR) constitute a small class of isolated 
neutron stars; four are currently well-established while one more 
source remains an unconfirmed candidate (Cline et al. 2000). As their name 
indicates, SGR emission is repetitive: during their active states each 
source emits hundreds of short (duration $\sim$ 0.1 s), intense (at 
$\sim$ super-Eddington luminosities) bursts of hard X-rays / soft 
gamma-rays at random intervals. Very rarely, (observed once so far 
from each of SGR~$0526-66$ and ~$1900+14$) they emit a giant flare, an 
event with $\sim$10$^4$ higher luminosity than the typical short 
events (Mazets et al. 1979, Hurley et al. 1999a, Feroci, et al. 2001). 
Thompson \& Duncan (1995; henceforth TD95) proposed that the SGR burst 
activity is associated with the strong magnetic field (B$\gtrsim$ 
10$^{14}$ G) of the neutron star. Their model, also known as the 
magnetar model, attributes SGR bursts to various levels of 
fracturing of the neutron star crust by the motion of the anchored 
lines of their strong magnetic fields (TD95, Thompson \& Duncan 2001). 

SGR sources were established as a new class in the mid 80s based on 
their bursting properties. The first X-ray counterpart of an SGR was 
discovered in 1993 with the Japanese satellite ASCA (Murakami et al. 
1994) after SGR~$1806-20$ triggered the Burst And Transient Source 
Experiment (BATSE) onboard the Compton Gamma Ray Observatory (CGRO) 
(Kouveliotou et al. 1994). All four SGRs are currently associated with 
persistent X-ray point sources with luminosities 10$^{34}$ $-$ 
10$^{35}$ ergs s$^{-1}$. Their energy spectra are nominally fit with a 
single power law with photon indices $2-3.2$, except in the case of 
SGR~$1900+14$, where a two component spectrum (blackbody of 
$kT=0.50(4)$ and power law with index $=2.1(3)$) has been established 
with BeppoSAX and Chandra observations (Woods et al. 1999a, 
Kouveliotou et al. 2001). 

Observations of SGR~$1806-20$ with the Rossi X-Ray Timing Explorer (RXTE) 
Proportional Counter Array (PCA) on November 1996 
led to the discovery of the first SGR spin period of 7.47 s. Further 
RXTE/PCA observations established that the source exhibited a very 
rapid spin-down rate of 8.1 $\times$ 10$^{-11}$ s s$^{-1}$, providing 
the first direct measurement of the neutron star magnetic field of $\sim$ 
2$\times$ 10$^{14}$ G\footnotemark, 
\footnotetext{Assuming that the star slows down via 
magnetic dipole radiation, B$_{\rm d}$ $\propto$ $\sqrt{{\rm P}\dot{\rm P}}$.}
in agreement with the magnetar model of TD95 (Kouveliotou et al.~1998a). 
ASCA observations of SGR~$1900+14$ in 1998 
led to the discovery of a spin period for that source (Hurley et al. 
1999b) and to the detection of its rapid ($1.1\times10^{-10}$ s 
s$^{-1}$) spin down rate (Kouveliotou et al. 1999). Of the two 
remaining SGRs, SGR~$0526-66$ has been associated with an 8 s spin 
period, originally seen as intensity modulation in the decaying tail 
of the giant flare it emitted in 5 March 1979 (Mazets et al. 1979). 
A pair of recent Chandra observations of this source during quiescence
spaced $\sim$ 1 year apart should establish the periodicity and spin
down rate of SGR~$0526-66$ (Kulkarni et al. 2002). 
The last source, SGR~$1627-41$, is the dimmest of them all and no 
spin modulation has been detected in its persistent X-ray flux
(Kouveliotou et al. 1998b). 

In 1998 we started a monitoring campaign with RXTE/PCA for the two SGR 
sources with clearly established spin and spin down rates. Our 
observations so far have shown that these spin down rates vary 
significantly, sometimes by a factor of four, and that they 
exhibit a very high level of timing noise (Woods et al. 2000, 2002). 
Earlier results have also indicated a relationship between the pulse 
profile complexity and the activity history of the source (Woods et 
al. 2001).  In this paper we construct the first detailed history 
of the pulse profile evolution of SGR~$1806-20$ and SGR~$1900+14$ 
spanning the last $\sim$ 5 years, and we examine their evolution 
with energy. We quantify the profile changes by estimating 
the power in their respective Fourier harmonics. In Section 2, we 
describe our observations and how we deal with the X-ray background 
component in our data. In Section 3, we describe the methodology of 
our data analysis and present the results; we discuss the implications 
of our results in Section 4. 

\section{Observations} 

The RXTE/PCA has 5 identical Proportional Counter Units (PCUs) with a 
total effective area of $\sim$ 6500 cm$^{2}$, a field of view of 
$\sim1^{\circ}$ FWHM, and is sensitive to photon energies between $2-60$ 
keV. Also onboard RXTE, the High Energy X-ray Timing Experiment 
(HEXTE) consists of two clusters of NaI/CSI scintillation detectors 
sensitive to photon energies of $15-250$ keV. Although we searched the 
HEXTE data, we have not detected any pulsations at the SGR 
frequencies; therefore, here we present only results obtained with the 
RXTE/PCA. 

During our ongoing monitoring campaign and also during occasional 
target of opportunity observations when burst activity was detected, 
we have observed SGR~$1806-20$ and SGR~$1900+14$ with RXTE over 2 Ms. 
In Table 1 we list the observation epochs, their date ranges, number 
of observation and total (on source) exposure times of each epoch for 
SGR~$1806-20$. Similarly, in Table 2 we list those for SGR~$1900+14$ 
with an extra column where we report the occasional presence of 
4U~$1907+07$ and XTE~J$1906+09$ in the PCA field of view. 

For each observation we used the {\it seextract} utility of FTOOLS 
5.0, to generate the source intensity light curves from the event mode 
PCA data with 0.125 s time resolution in 3 energy bands: $2-5$ keV, 
$5-10$ keV and $10-20$ keV. We filtered out the times of data 
anomalies, such as times of large pointing offsets and times of SGR 
burst activity, using the appropriate housekeeping data to obtain the 
persistent SGR emission plus the background. As the PCA is currently 
operating with varying number of PCUs, we normalized the rates of each 
time bin to the number of active PCUs. Finally we corrected the photon 
arrival times to the solar system barycenter using {\it fxbary} for 
observations prior to 2000 December 31 and {\it faxbary} for the later 
ones.	

\subsection{Background Issues} 

Estimating the X-ray background emission with the PCA is not a trivial 
problem. We had to account for various background components included 
in the observed emission simply due to the relatively large field of 
view of the PCA. Since both SGRs are located in the galactic ridge 
region, there are also inevitably other sources occasionally active in 
the same field. During the vast majority of these observations the count 
rates of both SGRs were very low in comparison to the total (source +
background) count rate; namely in 1996 epoch of SGR 1806$-$20 the average
total rates in $2-5$ keV, $5-10$ keV and $10-20$ keV bands were 44.32, 51.76
and 47.36 counts s$^{-1}$ PCU$^{-1}$, respectively, while the expected source
count rates (estimated by employing WebPIMMS with a interstellar medium 
attenuated [N$_{H}$ = 6.1 $\times$ 10$^{22}$ cm$^{-2}$] power law model 
[$\Gamma$ = 1.97], Mereghetti et al. [2000]) were 0.35, 0.57 and 0.31 
counts s$^{-1}$ 
PCU$^{-1}$, respectively. Similarly the average observed count rates for
SGR 1900+14 were 39.36, 49.52 and 50.32 counts s$^{-1}$ PCU$^{-1}$ in three
energy bands, whereas the expected source rates were only 0.27, 0.35 and 0.19
counts s$^{-1}$ PCU$^{-1}$ (using the spectral model of a blackbody 
[kT$_{\rm BB}$=0.5 keV] plus a power law [$\Gamma$ = 2.1], both attenuated
by the interstellar absorption [N$_{H}$ = 2.3 $\times$ 10$^{22}$ cm$^{-2}$],
Kouveliotou et al. [2001]). 
Below we describe the different components that 
contribute to the background of each source and our efforts to 
determine their values. Nonetheless, we are unable to estimate the 
background level with enough accuracy to determine the source count rates. 

~~~~{\it (i) Instrumental Background:~}~ This is due to events created 
by the particles in the vicinity of the instrument. For each 
observation, we have created a PCA background event data file using 
the faint background models (provided by the PCA instrument team for 
each epoch). We then extracted the background light curves in all 3 
energy intervals and processed them as described in Section 2. We then 
determined the average background rate per PCU per epoch. 

~~~~{\it (ii) Galactic Ridge Contribution:~}~ There have been numerous 
attempts to model the diffuse emission of this region. The most recent 
and extensive one is by Valinia \& Marshall (1998) using RXTE 
measurements. We reproduced the diffuse galactic ridge spectrum
by employing their estimated parameters of a two-component 
thermal plasma plus power law model for their R1 region (within which 
both SGRs are located) and the PCA response matrix  (with all 5 PCUs 
operating). We then estimated the count rates (per PCU) in each of our 
energy bands for each epoch. 

~~~~{\it (iii) Point X-ray Background Sources:~}~ There are two known 
point X-ray sources in the vicinity of SGR~$1900+14$; 4U~$1907+09$ (a
persistent 
X-ray pulsar with spin period of $\sim$ 440 s, located 0.51$^{\circ}$ 
away from the SGR) and XTE~$J1906+09$ (a transient X-ray pulsar with 
89 s spin period at 0.28$^{\circ}$ away). For the former source, we 
have used the average spectral model parameters and flux given by 
Roberts et al. (2001) to estimate its count rate. We then used the PCA 
collimator response at the pointing offset to this source during 
each observation to estimate the rates from 4U~$1907+09$ that would be 
present during our observations. For the latter source, Wilson et al. 
(2002) have performed an extensive search for periods of outburst 
activity, using the same data set presented here up until the end of 
2000. They determined two active episodes, one each in 1996 and 1998. 
We then determined the expected rates from XTE~$J1906+09$ by using 
their best model parameter estimates for each of the outburst periods, 
following the same procedure as for the former source. 

After subtracting the background contribution estimated by taking 
account all the above described components, we constructed the $2-5$ 
keV and $5-10$ keV pulse profiles for each source. We then compared 
our results with those obtained (within the same energy bands) from 
contemporaneous measurements with imaging instruments (BeppoSAX/NFI 
observations of SGR~$1900+14$ in September 1998 and Chandra/ACIS$-$S 
observations of SGR~$1806-20$ in August 2000). Even though the pulse 
profiles resembled each other significantly, the RMS pulse fractions 
obtained with the PCA measurements were smaller than those of BeppoSAX 
or Chandra by a factor of $\sim$ 3. (Here we define the pulse fraction 
as $PF_{\rm RMS}=
\sqrt{\sum_{i=1}^{N}[R_i-R_{ave}]^2/R_{min}^2} / N$, 
where R represents the count rate in each phase bin). These results 
indicate that there is still a significant background contribution in 
our data, or alternatively, that despite all our efforts we still 
severely underestimate our background. Although we cannot accurately 
determine our source of error, we believe that it is mostly due to 
underestimating the galactic ridge background contribution. We will 
refrain, therefore, in the following from calculating pulse fractions 
or hardness ratios (ratios of different energy fluences) that are 
based on background-subtracted data and will use the 
background-independent method described in Section 3 to draw physical 
conclusions on the pulse profile properties. 

\section{Pulse Profiles} 

In several of our observations the sources were offset from the center 
of the PCA field of view to either reduce the likelihood of saturation 
from bursts (when they were 
very active) or to reduce occasional contributions from other 
transients. We have taken these pointing offsets into account by applying 
the appropriate collimator response corrections for all datasets.  
We then created 
pulse profile histories in three energy bands ($2-5$, $5-10$ and $10-20$ 
keV) for each source using phase-connected pulse ephemerides 
reported elsewhere (Woods et al. 1999b, 2000, 2002). 

We quantify the pulse profile shapes (and changes) with the powers of 
their Fourier harmonics: for each source, we Fourier transformed each 
profile and calculated the normalized Fourier Powers (FP) of the first 5 
harmonics as follows. We first estimated the powers as  
P$_{\rm k}$ = 2(a$_{\rm k}^2$~+~b$_{\rm k}^2$)/($\sigma^2_{\rm a_{\rm k}}~+~
\sigma^2_{\rm b_{\rm k}})$, where a$_{\rm k}$ and b$_{\rm k}$ 
are the coefficients of the sine and cosine terms of the Fourier series, 
respectively, and $\sigma_{\rm a_{\rm k}}$ and $\sigma_{\rm b_{\rm k}}$
are the standard deviations of these coefficients. This is equivalent to the
Leahy normalization standardly used in X-ray astronomy.
Using the formalism described by Groth (1975) with the estimated power values, 
we measured the median and 68 \% significance level of the Groth distribution.
We then corrected the estimated powers for the binning of the pulse profile, 
which is a prominent effect on the powers of higher harmonics, using
equation 2.19 of van der Klis (1989). The resulting
powers of each profile were then normalized by the total power.   

We have applied the following procedure to obtain a quantitative measure of 
the level of pulsed intensity of the source during each observation. We 
estimated the average count rate, $R_{\rm ave}$, of each profile and then 
calculated the rms pulsed count rate, $PCR_{\rm rms}$, per source as 
$PCR_{\rm rms}=\sqrt{\sum_{i=1}^{N}[(R_{\rm i}-R_{\rm 
ave})^2 - (\Delta R_{\rm i})^2] / N}$, where R$_{\rm i}$ is the
count rate in each phase bin, $\Delta R_{\rm i}$ are the associated Poisson
statistical errors and N is the number of phase bins (which is 20). 
Typically, $\sum_{i=1}^{N}(\Delta R_{\rm i})^2$ is about 23, 9 and 37 $\%$ of
$\sum_{i=1}^{N}(R_{\rm i}-R_{\rm ave})^2$ for the $2-5$, $5-10$ and $10-20$ keV
pulse profiles of SGR~$1806-20$. Note that the above percentage can be as
high as 81 $\%$ if the pulsation is very weak. 
The rms value of the pulsed count rate is a background-immune measure and 
provides a reliable indicator of the pulsed intensity.

\subsection{SGR~$1806-20$} 

We exhibit in the three upper panels of Figure 1 the pulse profile 
changes over time and energy of the source. Note that the rates in the 
profiles are in arbitrary units on account of the background contamination 
as explained in Section 2. The bottom panel displays 
the quantitative evolution of the harmonic contents in its Fourier 
spectrum. At all epochs, we find that most of the power is in the first
harmonic of the Fourier spectrum. 

\begin{figure}
\vspace{0.3in} 
\plotone{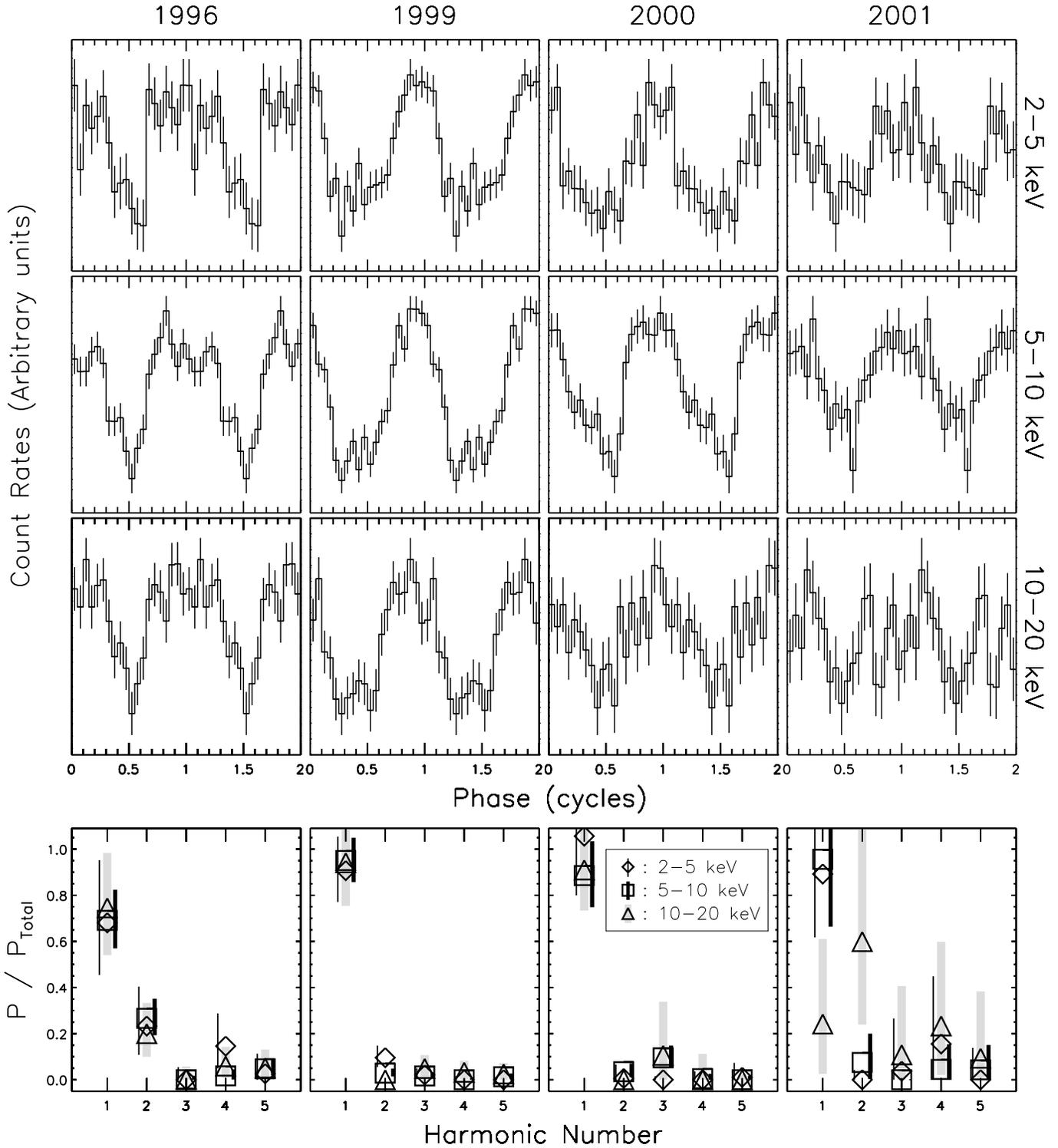}
\vspace{0.1in} 
\caption{({\it upper 3 panels}) Pulse profiles of SGR 1806-20 as observed in 
1996, 1999, 2000 and 2001 (from left to right) in 2$-$5 keV, 5$-$10 keV 
and 10$-$20 keV energy bands. The count rates are in arbitrary units as
explained in the text; ({\it bottom panel}) Normalized power of the pulse 
profile Fourier harmonics vs associated harmonic number for each profile.
The error bars associated with the 2$-$5 keV and 5$-$10 keV values are shifted
to the left and right, respectively, for displaying purposes.}
\end{figure}

As reported in Woods et al. (2002), observations between 2000  
July 4 to September 3 could not be phase 
connected. For these data we employed a phase-aligning technique to 
obtain the overall pulse shape. We first created a template pulse 
profile using 55 days of a phase-connected timing solution (data 
between 2000 May 2 and June 25). We then folded each of our non-phase 
connected data sets at the spin frequency measured for these data sets 
(Table 3 in Woods et al. 2002). Next, we determined the phase offset 
of each profile with respect to this template by fitting the first four
Fourier coefficients of the folded profile to the same
coefficients of the template as described in Finger et al. (1999).
Then, finally we aligned the profiles, and obtained the average 
pulse shape for the 2000 epoch of the source. The relative alignment errors
were small (less than 5 $\%$) so that this process should not introduce any
significant smearing of the pulse profile.

The pulse harmonic content before and during the 1996 
source activation is distributed into two harmonics.
We investigated the evolution of the pulse profile 
during our two week long RXTE observations in 1996 (which is the last 
period where we observe a complex profile) by subdividing our data
into three sets of $\sim$ 45 ks each.
We find that all profiles are consistent with each other,
showing no major changes in complexity.
In 1999 and 2000, the pulse profiles are significantly sinusoidal and
resume a marginally more complex shape during late 2001.
As the source was barely active during 
this latter period, however, the pulsed signal is weaker and almost 
undetectable above 10 keV. In general, the pulse profiles are 
relatively wide, covering $\sim$80\% of the phase cycle, except in the 
1999 observations where it becomes narrower covering $\sim50\%$ of the 
cycle. 

\begin{figure}
\plotone{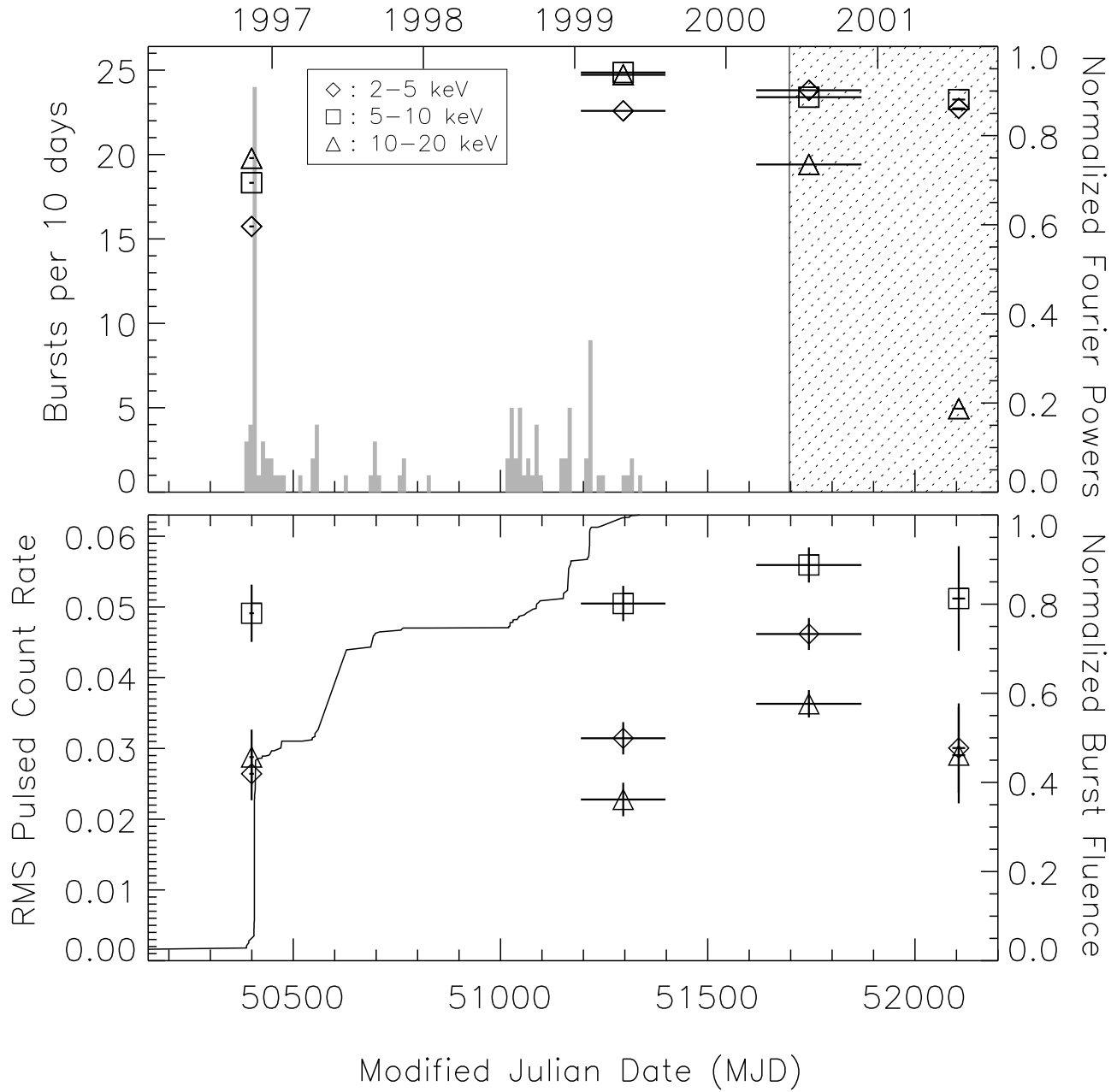}
\caption{({\it upper panel}) History of the power in the fundamental 
(3 energy bands) for SGR 1806-20 along with the burst activity 
history of this source as seen with BATSE. The horizontal bars at the
Fourier power points denote the time spans of each observing cycle. 
The dotted region starts after the reentry of CGRO. ({\it bottom panel}) 
The history of the rms pulsed count rate in each energy range 
during the same period. The solid line is the cumulative normalized
burst fluences detected with BATSE.} 
\end{figure} 

To investigate the dependence of the pulse shape on the source 
activity we plot in the upper panel of Figure 2 the source burst 
activity (grey histogram) detected with BATSE covering the period of 
our RXTE monitoring program until the reentry of CGRO on June 2000. 
The data points in the plot are the values of the normalized Fourier 
power in the first harmonic {\it only} (in three energy bands). The 
power increases during 1999 and reduces down to the 1996 level in 
2001. The bottom panel in Figure 2 exhibits the rms pulsed count rate 
of the source in each energy range overplotted on the cumulative energy 
distribution of all bursts from the source detected with BATSE since 1996. 
We note that the count rate (and thereby the source pulsed flux intensity) 
remains fairly constant during 
our observations and that the pulsed rates between $5-10$ keV are always 
larger than those in each remaining energy band. 
There is no measurable variability in the rms pulsed count rate with burst
activity. There is, however, intense burst activity between our pulse shape
measurements in 1996 and 1999 which may be the cause for this pulse shape
change (see also \S 4).

\subsection{SGR~$1900+14$} 

We have performed a similar analysis for SGR~$1900+14$. Figures 3 
through 5 exhibit the pulse shape and its harmonic content evolution 
over a series of quiescent and active episodes of the source. 

In 1996, the source was in quiescence with no burst activity detected. 
A three-peak pulse profile is seen in all three energy bands, with all 
peaks aligning in phase (Figure 3, left column). As we can 
also see in the FPs plot, the majority of power is in the third 
harmonic of each profile. The distinct, rapid, large modulation in the 
10$-$20 keV has not been seen with similar significance in any other 
epoch. Even though the source entered a burst active episode in the 
end of May 1998 (emitting numerous, short SGR bursts), the shape of 
its pulse profile resembles the one of the 1996 observations  (Figure 
3, middle column). 

\begin{figure}
\vspace{0.25in} 
\plotone{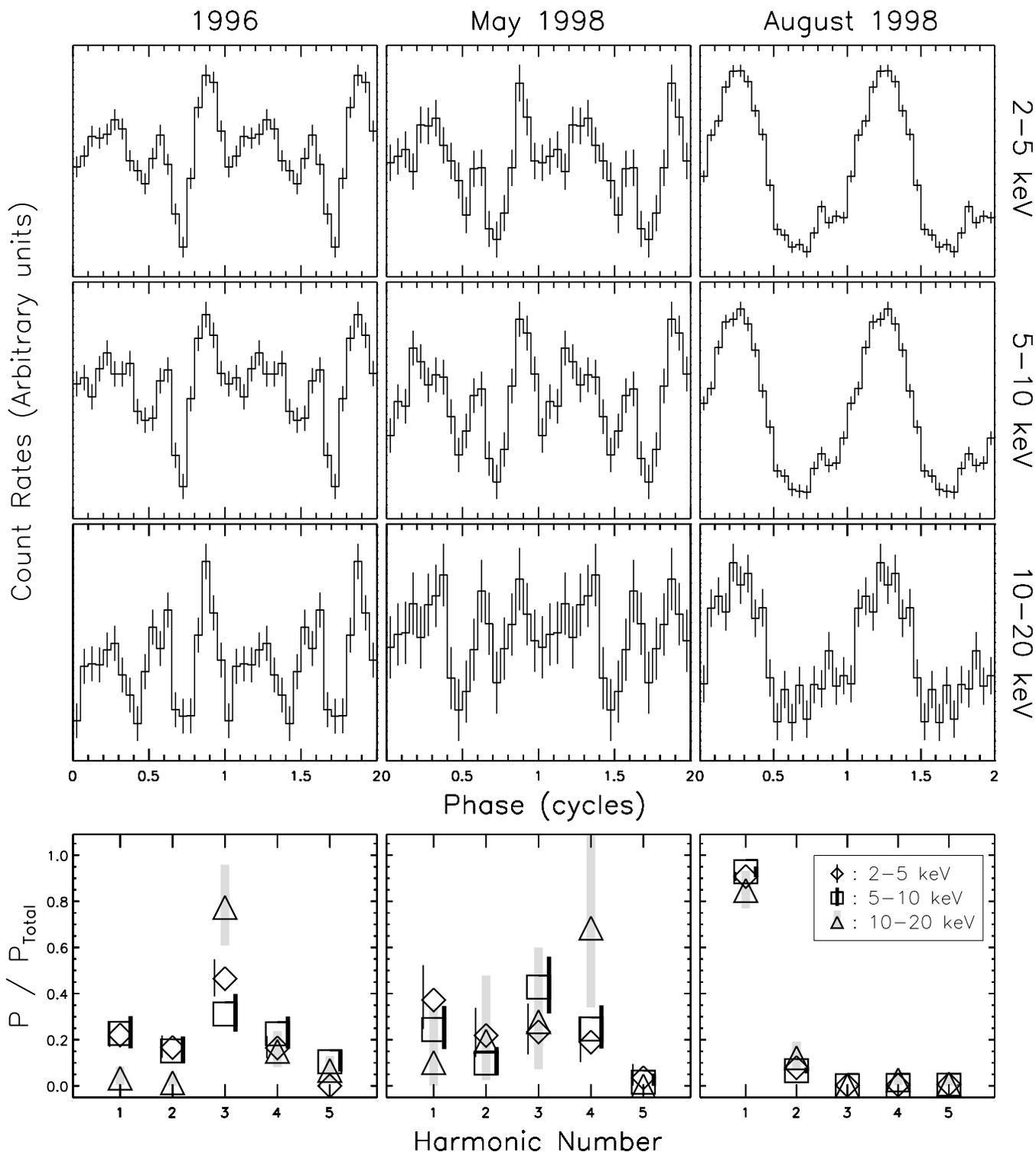}
\vspace{-0.01in} 
\caption{({\it upper 3 panels}) Pulse profiles of SGR 1900+14 as 
observed in 
1996, May 1998, and August 1998 (from left to right) in 2$-$5 keV, 5$-$10 keV 
and 10$-$20 keV energy bands. The count rates are in arbitrary units as
explained in the text; ({\it bottom panel}) Normalized power of the pulse 
profile Fourier harmonics vs associated harmonic number for each profile.
The error bars associated with the 2$-$5 keV and 5$-$10 keV values are shifted
to the left and right, respectively, for displaying purposes.} 
\end{figure} 

On 27 August 1998, however, SGR~$1900+14$ emitted a `giant flare' as 
described in the Introduction. In the tail emission of this burst, 
which was strongly modulated at the spin period of the neutron star, 
Feroci et al. (2001) observed a highly significant four-peaked pulse 
profile (between t+40 s and t+90 s, where t is the trigger time of 
flare); the profile became almost sinusoidal at $\sim$ t+300 s. Our 
August 1998 pulse profile here covers four days starting the day after 
the burst. The pulse profile is also simplified to a broad main pulse 
plus a secondary peak (or shoulder) at $\phi \sim$0.8  (Figure 3, right 
column). The main pulse covers almost half the spin cycle. 

\begin{figure}
\vspace{0.25in} 
\plotone{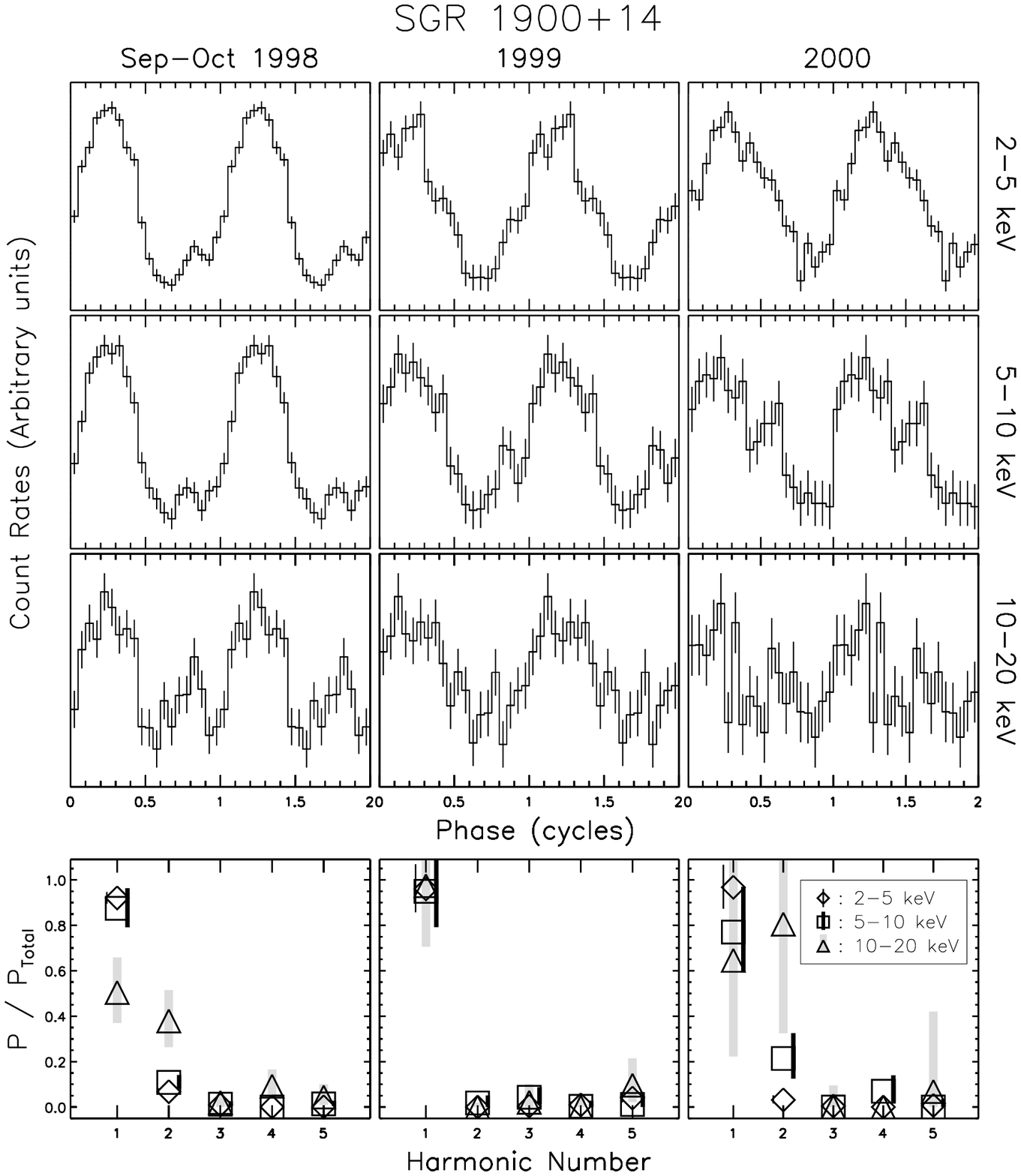}
\vspace{-0.01in} 
\caption{({\it upper 3 panels}) Pulse profiles of SGR 1900+14 
as observed in
September-October 1998, 1999 and 2000 (from left to right) in 2$-$5 keV,
5$-$10 keV and 10$-$20 keV energy bands. The count rates are in arbitrary
units as explained in the text; ({\it bottom panel}) Normalized power of the
pulse profile Fourier harmonics vs associated harmonic number for each profile.
The error bars associated with the 2$-$5 keV and 5$-$10 keV values are shifted
to the left and right, respectively, for displaying purposes.}
\end{figure} 

Over the next 2 months after the giant event, the 2$-$5 keV band pulse 
profile remained almost identical to that of the very short epoch soon 
after the flare (Figure 4, left column). In the 5$-$10 keV profile, 
the main pulse structure did not change but there is some evidence 
that the secondary peak feature appears earlier in phase than in the 
2$-$5 keV profile. The secondary peak structure is more significant 
between $10-20$ keV during this epoch. The main pulses in 10$-$20 keV 
during both August 1998 and September-October 1998 cover a slightly 
shorter phase (compared to the lower energy bands) and display sharp 
rise and fall times. 
	  
SGR~$1900+14$ observations between 1999 January 25 and July 27 could 
not be phase connected (Woods et al. 2002). Similar to SGR~$1806-20$, 
we used the pulse profile of a 14 day segment of phase connected data 
early in  January 1999 as our template. With this template and using 
individual pulse frequency measurements for non-phase connected 
observations as given by Woods et al. (2002) we employed the same 
phase-aligning procedure to get the average pulse profile of 
SGR~$1900+14$ in 1999 (Figure 4, middle column). The 2$-$5 keV and 
5$-$10 keV profiles are described by a broad single pulse covering 
almost 75\% of the  phase cycle; there is no evidence of a secondary 
peak in the FP in all bands. 

In 2000, the pulse profiles in the 2$-$5 keV and 5$-$10 keV bands are 
dominated by a single pulse with $\sim$75\% of phase coverage (Figure 
4, right panel). Both profiles show a  sharp rise followed by a gradual 
fall. The high power in the second harmonic of the 2$-$5 keV band is 
due to the emission enhancement around  $\phi \sim$0.6. In this epoch, 
only weak pulsations were seen in the 10$-$20 keV energy band. 

\begin{figure}
\vspace{0.25in} 
\plotone{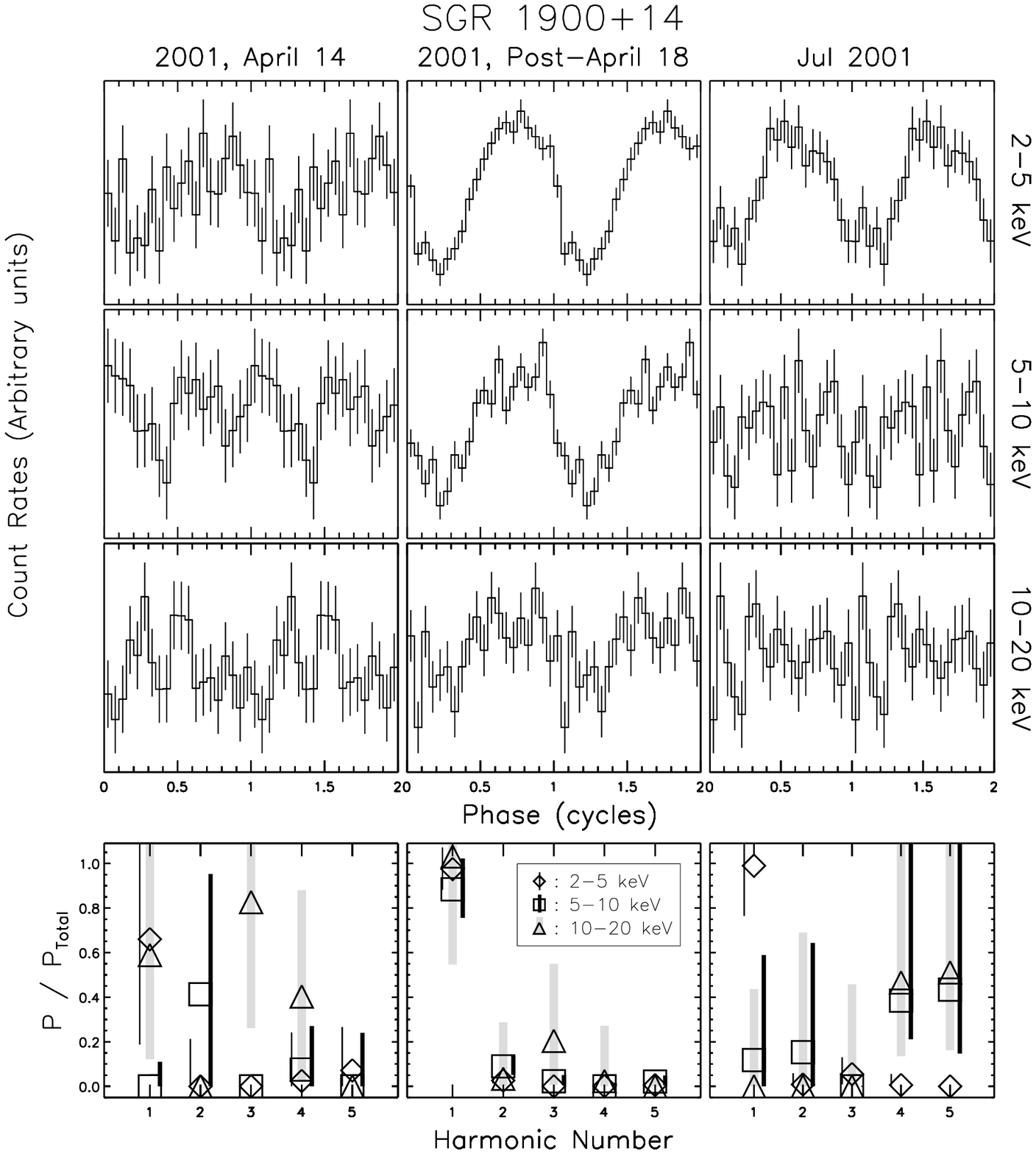}
\vspace{-0.01in} 
\caption{({\it upper 3 panels}) Pulse profiles of SGR 1900+14 
as observed on 14 
April 2001, April-May 2001 and June-July 2001 (from left to right) in 2$-$5 keV,
5$-$10 keV and 10$-$20 keV energy bands. The count rates are in arbitrary
units as explained in the text; ({\it bottom panel}) Normalized power of the
pulse profile Fourier harmonics vs associated harmonic number for each profile.
The error bars associated with the 2$-$5 keV and 5$-$10 keV values are shifted
to the left and right, respectively, for displaying purposes.}
\end{figure} 

Our ongoing SGR monitoring campaign with RXTE revealed that 
SGR~$1900+14$ was in quiescence at least over the last 8 months in 
1999 and throughout 2000. The source entered a new episode of burst 
activity starting with an intermediate intensity SGR flare detected on 
18 April 2001 (Guidorzi et al. 2001). This event resembled the August 27 
giant flare in some respects but was weaker energetically. Only four 
days prior to the reactivation, we performed a regular RXTE monitoring 
observation. We show the pulse profiles obtained from this 
observations in Figure 5 (left column). The signal is extremely  weak 
and does not allow detection of pulsations in any of the energy bands
except the 2$-$5 keV.  

The reactivation of the source initiated a series of ToO observations 
whose timing results are presented in Woods et al. (2002). 
Using the spin ephemeris we determine 
the profiles shown in Figure 5 (middle column). Even though the 
change in pulse shape after the April 18 intermediate flare is not as 
dramatic as the one after the August 27 giant flare, 
the pulse profile becomes significantly more sinusoidal following
this event. In the 2$-$5 keV profile, we observe one of the 
broadest single pulse structures ever seen from this source. In July 
2001, the long term monitoring campaign observations with RXTE were 
resumed. We observe significant pulsations only in 2$-$5 keV profile which
is being well described by a single pulse (Figure 5, right column).

\begin{figure}
\plotone{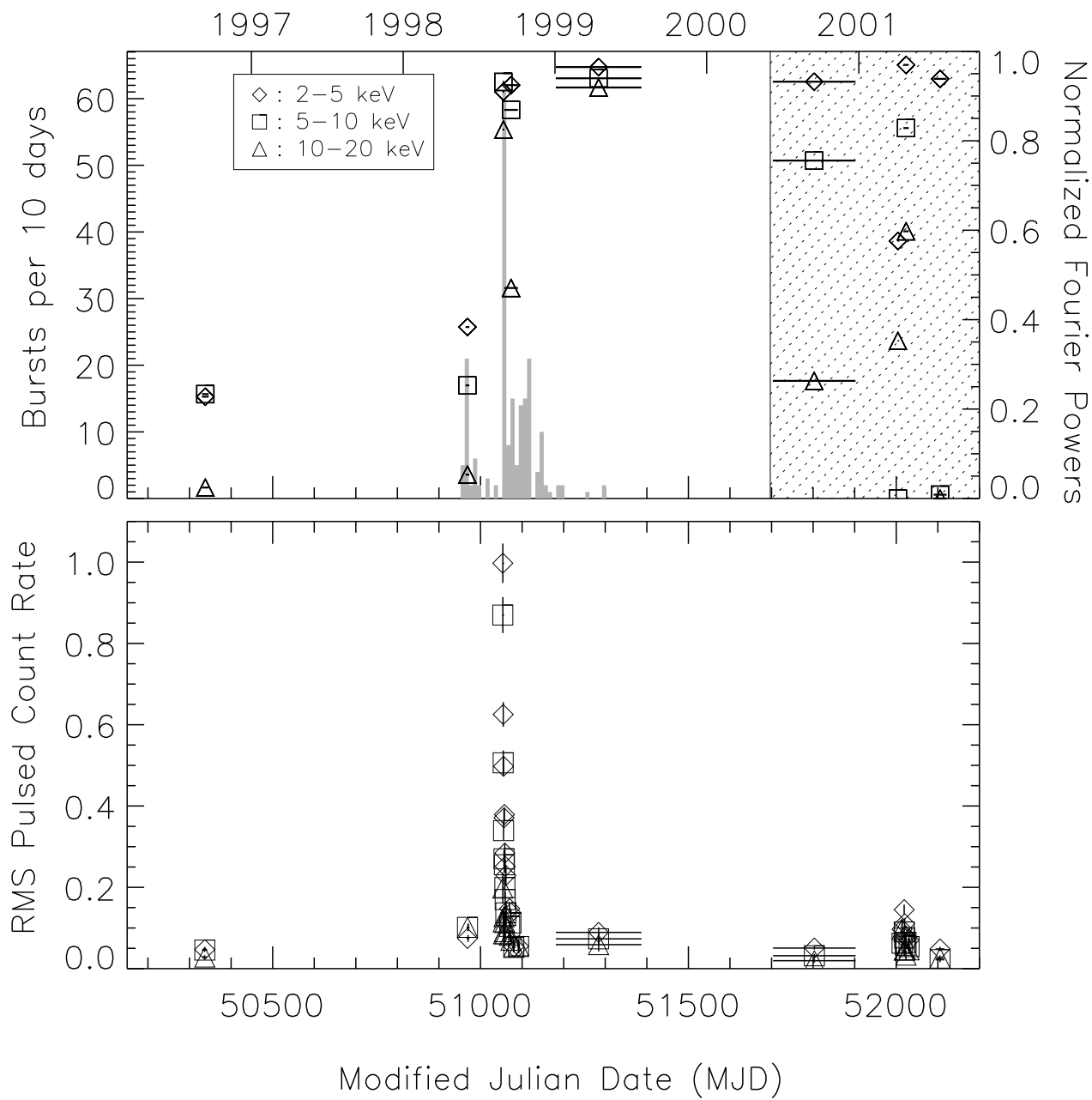}
\caption{({\it upper panel}) History of the power in the fundamental 
(3 energy bands) for SGR 1900+14 along with the burst activity 
history of this source as seen with BATSE. The horizontal bars at the
Fourier power points denote the time spans of each observing cycle. 
The dotted region starts after the reentry of CGRO. ({\it bottom panel}) 
The history of the rms pulsed count rate in each energy range 
during the same period.} 
\end{figure} 

Next, we plot the burst activity of SGR 1900+14 (grey 
histogram) versus the power in the first harmonic (Figure 6 {\it upper 
panel}), as was done for SGR 1806$-$20. We notice that the pulse in 
each energy band is strongly sinusoidal 
after the August 27 flare, while complex before, consistent with previous 
results (e.g. Woods et al. 2001). We do not have good 
enough statistics to determine the pulse profile immediately before the 
April 18 event, however, there is a significant reduction in power
at the second harmonic from the year 2000 to post-April 18.
In the bottom panel of Figure 6 we plot the PCR of the source 
over time. We clearly see that the pulsed count rate follows the 
quiescent source flux trend reported by Woods et al. (2001). The 
pulsed count rate increases significantly after the August and April 
large and intermediate flares, respectively; it resumes its background 
value in all other intervals. We expand on these results in the next 
section.

\section{Discussion} 

Earlier results (Woods et al. 2001) have shown that the persistent 
flux of SGR~$1900+14$ decayed as a power law for almost 40 days 
following the August 27 flare (F $\propto$ t$^{-0.71}$). We have very 
good statistics in our data to study the pulsed count rates (PCR) 
measured during the decay of this event. In Figure 7 ({\it top panel}) 
we expand the tail of the event into ten individual measurements of 
PCRs in 3 energy bands over a 20-day interval. The horizontal lines in 
the plot denote the PCR levels in the 2$-$5 keV (dashed line), 5$-$10 
keV (dot-dot-dot-dashed line) and 10$-$20 keV (dot-dashed line) during 
the {\it quiescent} state of the source in 1996. Following the giant 
flare, the values of the PCRs increased by a factor of $\sim$ 21, 19, 
and 7, in the 2$-$5 keV (PCR$_{1}$), 5$-$10 keV (PCR$_{2}$), and 
10$-$20 keV (PCR$_{3}$) bands, respectively, over their quiescent 
levels. The middle and bottom panels of Figure 7 exhibit the  
ratios of PCR$_{1}$ to PCR$_{2}$ and PCR$_{1}$ to PCR$_{3}$, 
respectively. These trends can be interpreted as reflecting the 
spectral variations of the pulsed emission during the decay of the 
flare. The PCR$_{1}$/PCR$_{2}$ values varied between 1.14 $\pm$ 0.03
and 1.67 $\pm$ 0.16, always significantly higher than the non-active
episode value of 1.03. There is statistically significant 
variability in this softness ratio as a function of time, but
the changes are relatively small in amplitude ($\lesssim$ 40$\%$).
The PCR$_{1}$/PCR$_{3}$ ratio ({\it the bottom panel}) shows a clear 
hardening of the spectrum after the first $\sim$ 3 days of the decaying 
tail of the August 27 flare. 

\begin{figure}
\plotone{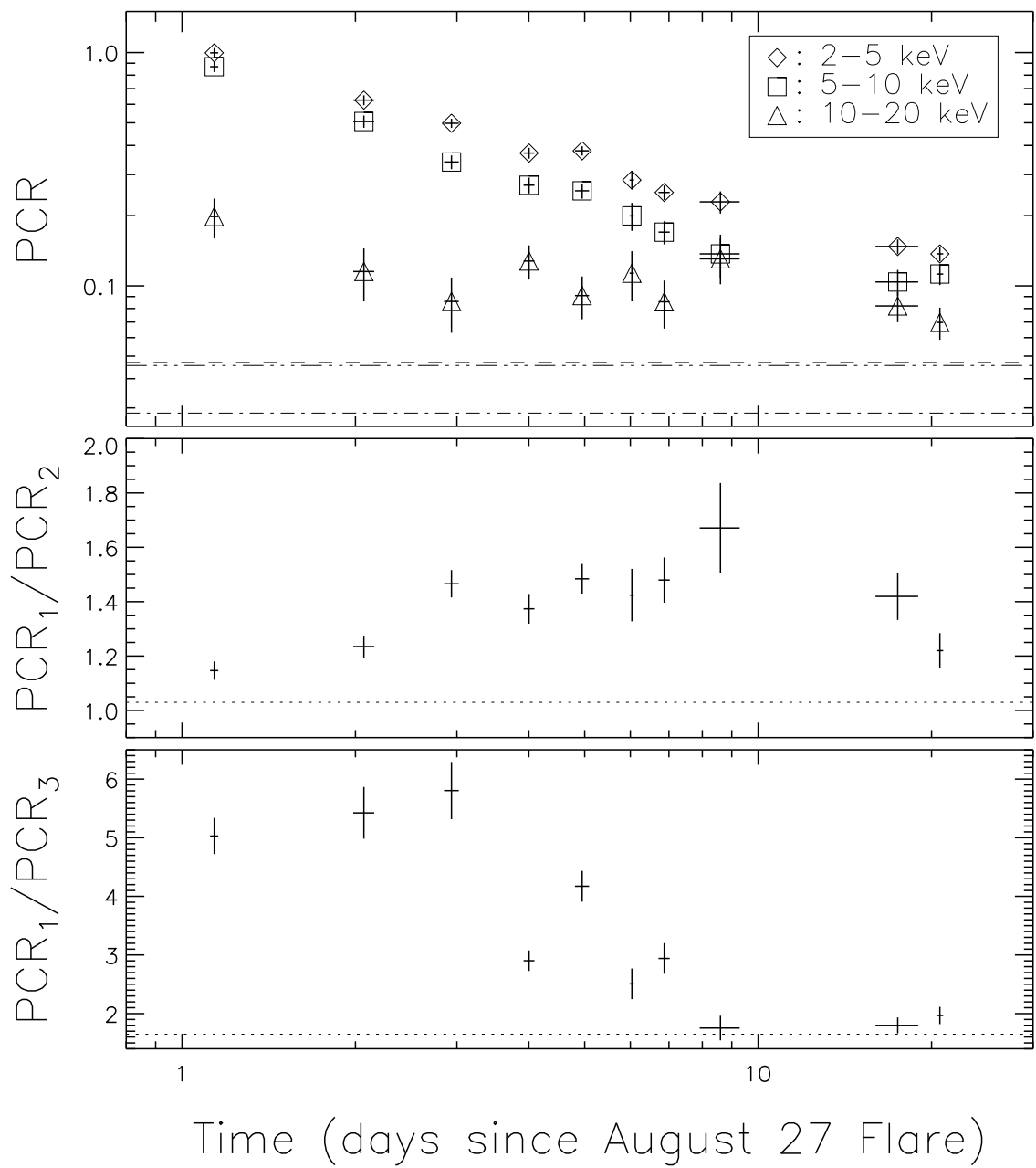}
\caption{({\it top panel}) Plot of pulsed count rate (PCR) 
decay in 3 energy bands following the 
August 27 flare. The corresponding energy of each symbol is presented in the 
legend. The horizontal lines are the PCR values (dashed line in 2$-$5 keV, 
dot-dot-dot-dashed line in 5$-$10 keV and dot-dashed line in 10$-$20 keV) 
in 1996 observations when there was no bursting activity. ({\it middle panel}) 
The ratio of PCR$_{1}$ to PCR$_{2}$ during the same time period. 
The vertical bars represent the errors associated with ratios and the 
horizontal bars show the observation time span over which PCRs are determined. 
The horizontal dotted line shows the PCR$_{1}$/PCR$_{2}$ ratio for the 
non-active episode. ({\it bottom panel}) The ratio of PCR$_{1}$ to PCR$_{3}$.} 
\end{figure} 

Figure 8 describes the decay behavior of the April 18, 2001 
intermediate flare from SGR~$1900+14$. The pulsed count rates here 
increased by a factor of 3.6, 2.6 and 2.2., for PCR$_{1}$, PCR$_{2}$, 
and PCR$_{3}$, respectively, again with respect to their quiescent 
values measured in 1996. As we also see in the middle panel, the decay 
of PCR$_{1}$ and PCR$_{2}$ are similar and their ratio is consistent 
with an average of 1.33, and there is marginal evidence of  
hardening starting 2 days after the event onset (bottom panel). Overall the 
soft-to-hard rate ratios are consistent with their quiescent values. 

\begin{figure}
\plotone{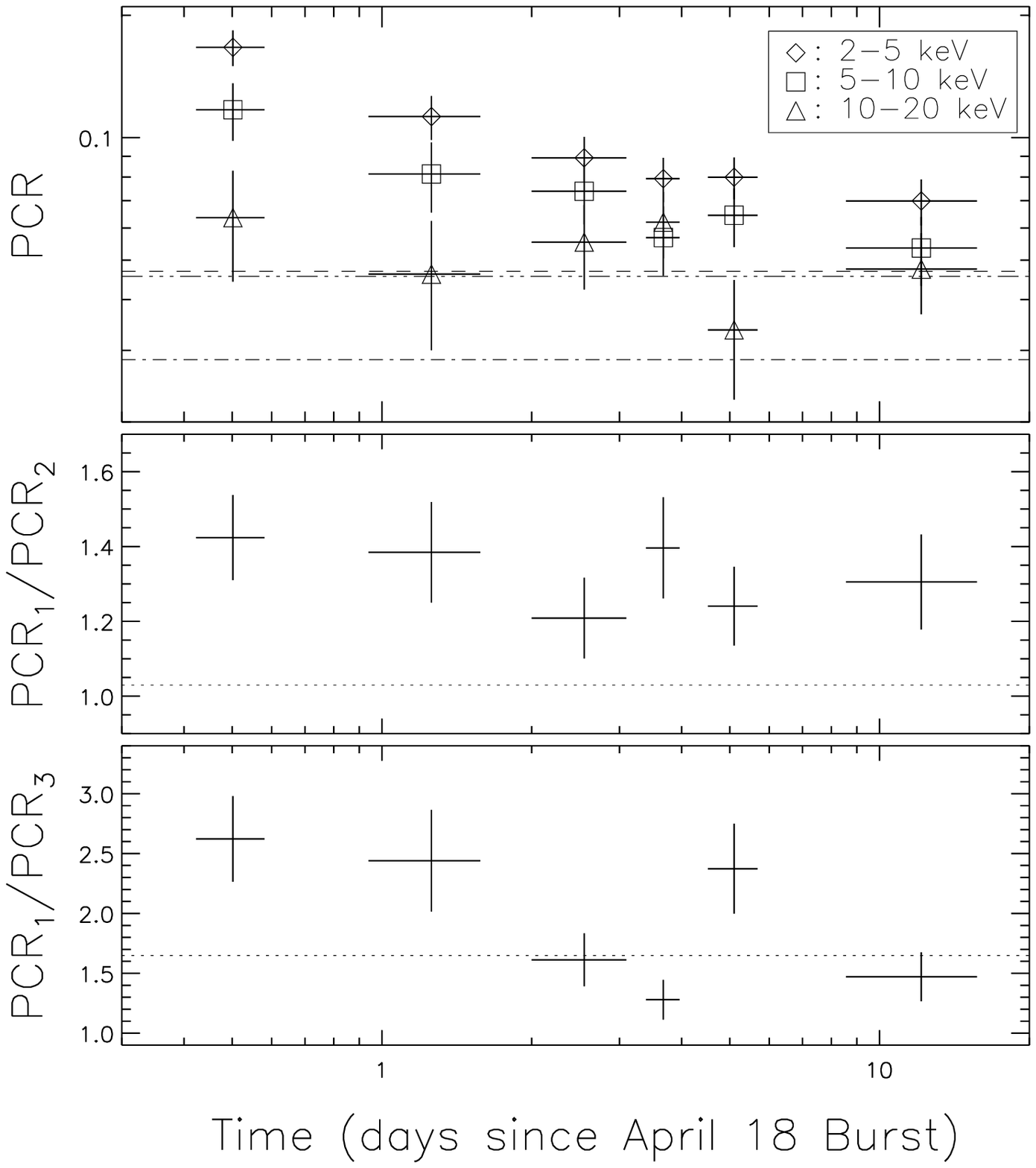}
\caption{({\it top panel}) Plot of pulsed count rate (PCR) 
decay in 3 energy bands following the 
April 18 burst. The corresponding energy of each symbol is presented in the 
legend. The horizontal lines are the PCR values (dashed line in 2$-$5 keV, 
dot-dot-dot-dashed line in 5$-$10 keV and dot-dashed line in 10$-$20 keV) 
in 1996 observations when there was no bursting activity. ({\it middle panel}) 
The ratio of PCR$_{1}$ to PCR$_{2}$ during the same time period. 
The vertical bars represent the errors associated with ratios and the 
horizontal bars show the observation time span over which PCRs are determined. 
The horizontal dotted line shows the PCR$_{1}$/PCR$_{2}$ ratio for the 
non-active episode. ({\it bottom panel}) The ratio of PCR$_{1}$ to PCR$_{3}$.} 
\end{figure} 

To compare the evolution with energy of the pulsed count rates in both 
sources, we plot in Figure 9 their PCR$_{1}$ vs PCR$_{2}$ (squares) 
and PCR$_{1}$ vs PCR$_{3}$ (triangles). The left panel of Figure 9 
exhibits all data points for SGR~1806$-$20 (filled symbols), together 
with measurements for SGR~1900$+$14, when the source was in the same 
(quiescent) state (open symbols). For SGR~1806$-$20 we find no 
correlation neither between PCR$_{1}$ and PCR$_{2}$ (filled squares), 
nor PCR$_{1}$ and PCR$_{3}$ (filled triangles).

The right panel of Figure 9 exhibits all our data points for 
SGR~$1900+14$, in both its quiescent and active states. We find no 
correlation between PCR$_{1}$ and PCR$_{3}$, 
but we find that PCR$_{1}$ and PCR$_{2}$ are strongly 
correlated with Spearman's rank order correlation coefficient,
$\rho$ = 0.99 and the probability of getting such a correlation from
a random data set, P$_c$ = 3.6 $\times$ 10$^{-8}$. We 
noticed, however, that there seems to be a break in the trend at very 
low values, so we reanalyzed these points separately. We find that a 
fit to PCR$_{1}>0.2$ yields a slope of 0.91 $\pm$ 0.05 ($\chi^{2}_\nu$ 
= 1.06); below this value, the fit has a slope of 0.69 $\pm$ 0.07 
($\chi^{2}_\nu$ = 0.93). The insert in Figure 9 (right panel) shows 
the extrapolation of both fits within the dashed line box. PCR$_{3}$ 
remains constant above PCR$_{1}>0.2$. We should note here, that all 
values outside the box belong to the decaying tail of August 27, 1998, 
while within the box, we have points from the same event as well as 
following the intermediate flare of April 18, 2001 and the quiescent source 
emission. 

\begin{figure}
\plotone{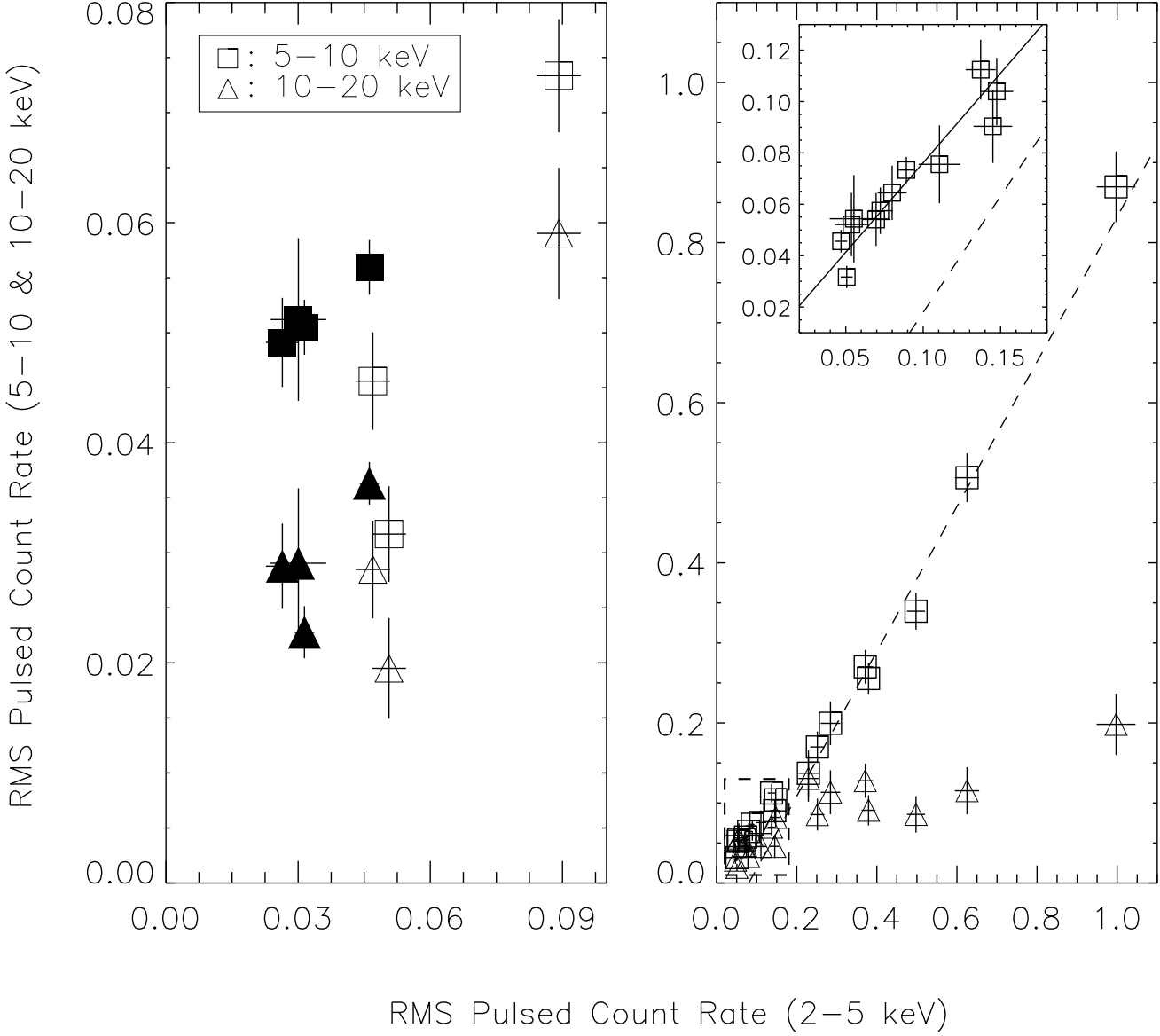}
\vspace{0.2in} 
\caption{({\it left panel}) Plot of PCR$_{1}$ vs PCR$_{2}$ (filled squares) and 
PCR$_{1}$ vs PCR$_{3}$ (filled triangles) for SGR 1806$-$20. The open 
symbols in this plot are those of SGR 1900+14 values measured during non active 
episodes. ({\it right plot}) Plot of PCR$_{1}$ vs PCR$_{2}$ (squares) and 
PCR$_{1}$ vs PCR$_{3}$ (triangles) for SGR 1900+14. 
The inset is the zoomed view of dashed box 
at the low end of PCR values. The solid line represents the best fit model to 
values with PCR$_{1}$ $>$ 0.2. The solid line within the inset is the 
best fit model to PCR$_{1}$ $<$ 0.2 values and the dashed line is the 
extrapolation of previous fit results.} 
\end{figure} 

SGR 1900+14 has shown dramatic pulse profile changes 
on a timescale of minutes associated with burst activity,
in particular during and after the giant flares. 
Unlike SGR 1900+14, changes in the SGR 1806$-$20 pulse profiles
are not drastic, but significant (c.f., those in 1996 and 1999
in Figure 1). Since our RXTE 
observations in 1996 ended before the peak of burst activity from this
source was recorded with BATSE, we looked at the energetics of bursts
seen with BATSE after the end of the RXTE data. The energy of BATSE events 
seen during the RXTE coverage 
comprise only $\sim$ 5$\%$ of the total energy (E$_{\rm TOT,BATSE}$) of all 
events seen with BATSE over $\sim$ 4 years (starting in 1996). 
Interestingly, bursts detected over the next $\sim$ 5 days following our 
last RXTE observations in 1996 contained $\sim$ 40$\%$ of
E$_{\rm TOT,BATSE}$. The later (1999) simplification of the pulse profile of 
SGR 1806$-$20 may, therefore, have its onset during 
this intense burst active phase in 1996, similar to the effect
of the 1998 August 27 giant flare in the pulse profile of SGR 1900+14.

\section{Conclusions}

We have performed a detailed analysis of the persistent X-ray pulsed
flux data from two Soft Gamma Repeaters, SGR 1900+14 and SGR 1806-20. In
both sources, we find a strong trend of their pulse shapes to transition
from complex to very smooth, following burst active episodes. Our
analysis of the decaying tail of the 27 August giant flare from SGR
1900+14, has shown that as the persistent source spectrum softened, the
pulse count rates in the 2-5 keV band varied in tandem with those of the
5-10 keV; the 10-20 keV pulsed count rates remained constant, albeit
slightly above their pre-flare values. Further, we observe a
change in the correlation slope of the pulsed count rates in the lower
energy ranges (2-5, and 5-10 keV) during quiescence or low activity of
the source, suggesting that the 2-5 keV photons are more efficiently
produced during non-burst active periods.

Recently, Thompson, Lyutikov \& Kulkarni (2002; TLK hereafter) 
have suggested a model, whereby the persistent
X-ray emission of a magnetar is due to currents generated when the
interior magnetic field twists up the external magnetic field. The
structure we observe in the pulse profile of SGRs could then be
attributed to a complicated distribution of currents and magnetic fields
close to the neutron star (higher multipoles) seen through a largely
transparent magnetosphere. However, the observed transition of complex
to smooth in the pulse profiles is not accompanied by spectral and
intensity changes, as would be expected in that case from a simplified
magnetic field (current) after the source returns to its quiescent flux
level. Alternatively, re-scattering of the X-rays either at $R>100$ km
by non-axisymmetric currents, or at ~100 km by plasma suspended against
gravity by the resonant e$^{-}$ scattering force, may be the cause of
the profile smoothness. The apparent difficulty to retain the suspended
plasma in the magnetosphere (it would be quickly drained to the neutron
star surface by even a modest electrical current), points towards the
re-scattering screen above 100 km as the most plausible of these two 
explanations for
the transition. The current then would in principle be generated by a
static twist (TLK 2002), consistent with the observed absence of a
direct correlation between the increase in the neutron star spin
(torque) with burst activity (Woods et al. 2002).

Concluding, we believe that our results provide evidence for an association
of the complex(smooth) profile of the SGR pulses with 
the re-scattering of X-rays above 100 km from the neutron star
surface, depending on whether the currents created by the static twist
described by TLK (2002) are non-axisymmetric(axisymmetric). The rapid
spindown observed in SGRs should then be coupled to the
smoothening of their pulse profiles; we are currently investigating
further correlations between all these SGR properties.

\acknowledgments 

We are grateful to Chris Thompson for many useful discussions and
to the anonymous referee for many helpful suggestions.  
We acknowledge support from NASA grants NAG5-6021, NAG5-7785 (E.G.);
the LTSA grant NAG 5-9350 (C.K. \& P.M.W.).

\newpage 

\begin{table}[!hp] 
\begin{center} 
\caption{{\it RXTE} PCA observations of SGR~1806$-$20. 
\label{tbl:1}} 
\vspace{11pt} 

\begin{tabular}{cccc} 
\hline 

Epoch	  & Date Ranges 	& No. of Obs. & Exposure    \\ 
	  & YY/MM/DD		& & (ks)      \\ 
\hline 
1996	  & 96/11/07$-$96/11/18 & 12	    & 132.4    \\ 
1999	  & 99/01/16$-$99/08/08 & 43	    & 297.9    \\ 
2000	  & 00/03/14$-$00/11/23 & 87	    & 254.4    \\ 
2001	  & 01/07/02$-$01/07/22 & 12	    & 37.5    \\ 
\hline 

\end{tabular} 
\end{center} 
\end{table} 

\begin{table}[!hp] 
\begin{center} 
\caption{{\it RXTE} PCA observations of SGR~1900+14. 
\label{tbl:2}} 
\vspace{11pt} 

\begin{tabular}{ccccc} 
\hline 

Epoch	      & Date Ranges	    & No. of Obs. & Exposure & 
Notes\tablenotemark{a} \\ 
	      & YY/MM/DD	    & & (ks)   & 
\\ 
\hline 
1996	      & 96/09/04$-$96/09/20 & 19	& 97.7   & 
J1906(A), 1907  \\ 
May 1998      & 98/05/31$-$98/06/09 & 5        & 43.5	& J1906(Q) 
\\ 
Aug 1998      & 98/08/28$-$98/08/31 & 10	& 33.2   & J1906(Q) 
\\ 
SepOct 1998   & 98/09/01$-$98/10/08 & 30	& 126.2   & J1906(A) 
\\ 
1999	      & 99/01/03$-$99/07/28 & 41	& 293.1   & 
J1906(Q), 1907 \\ 
2000	      & 00/06/08$-$00/12/24 & 93	& 435.6   & 
J1906(Q), 1907 \\ 
Pre-Apr 2001  & 01/04/14	    & 1        & 9.6   & 
J1906(N/A) \\ 
Post-Apr 2001 & 01/04/19$-$01/05/01 & 13	& 111.1   & 
J1906(N/A) \\ 
July 2001     & 01/07/04$-$01/07/25 & 22	& 81.5   & 
J1906(N/A) \\ 
\hline 

\end{tabular} 

\begin{flushleft} 
{\small 
a -- Presence of XTE J1906+09 or 4U 1907+09 within the PCA field of 
view. 
In paranthesis A, Q and N/A denote transient activity, quiescence and 
information 
not available, respectively.} 
\end{flushleft} 

\end{center} 
\end{table} 


\begin{references} 

\reference{Cli99} Cline, T.L., et al. 2000, \apj, 531, 407 
\reference{fe01} Feroci, M., Hurley, K., Duncan, R.C., \& Thompson C. 
2001, \apj, 549, 1021 
\reference{fin99} Finger, M., et al, 1999, \apj, 517, 449
\reference{Gro75} Groth, E.J., 1975, \apjs, 29, 285
\reference{gu01} Guidorzi, C. et al. 2001, GCN Circular 1041 
\reference{hur99} Hurley, K., et al. 1999a, \nat, 397, 41 
\reference{hur99b} Hurley, K., et al. 1999b, \apjl, 510, L111 
\reference{kou94} Kouveliotou, C., et al. 1994, \nat, 368, 125 
\reference{kou98a} Kouveliotou, C., et al. 1998a, \nat, 393, 235 
\reference{kou98b} Kouveliotou, C., et al. 1998b, IAU Circular 6944 
\reference{kou99} Kouveliotou, C., et al. 1999, \apjl, 510, L115 
\reference{kou01} Kouveliotou, C., et al. 2001, \apjl, 558, L47 
\reference{kul02} Kulkarni, S., et al. 2002, \nat, submitted 
\reference{maz79} Mazetz, E.P., et al. 1979, Nature, 282, 587 
\reference{mur94} Murakami, T. et al. 1994, \nat, 368, 127 
\reference{mer00} Mereghetti, S., et al. 2000, A \& A, 361, 240
\reference{rob01} Roberts, M.S.E., et al. 2001, \apj, 555, 967 
\reference{td95} Thompson, C. \& Duncan, R.C. 1995, \mnras, 275, 255 
\reference{td96} Thompson, C. \& Duncan, R.C. 2001, \apj, 561, 980 
\reference{TD01} Thompson, C. \& Lyutikov, M., \& Kulkarni, S.R., 2002, 
\apj, in press, astro-ph/0110677 
\reference{val98} Valinia, A. \& Marshall, F.E. 1998, \apj, 505, 134 
\reference{van89} van der Klis, M. 1989, in Timing Neutron Stars, eds. 
H. \"Ogelman \& E.P.J. van den Heuvel (Dordrecht: Kluwer), 27
\reference{wil01} Wilson, C.A., et al. 2001, \apj, 565, 1150
\reference{wo99a} Woods, P., et al. 1999a, \apjl, 519, L139 
\reference{wo99b} Woods, P., et al. 1999b, \apjl, 524, L55 
\reference{wo00} Woods, P., et al. 2000, \apjl, 535, L55 
\reference{wo01} Woods, P., et al. 2001, \apj, 552, 748 
\reference{wo01} Woods, P., et al. 2002, \apj, in press, astro-ph/0109361 

\end{references}
\end{document}